\documentclass[runningheads]{llncs}
\usepackage[T1]{fontenc}
\usepackage{graphicx}
\usepackage[colorlinks=true,linkcolor=blue,citecolor=blue,urlcolor=blue,]{hyperref}
\usepackage{algorithmic} 
\usepackage{algorithm} 
\usepackage{multirow}
\usepackage{booktabs}
\usepackage{cite}
\usepackage{color}

\begin{document}
\title{Natural Backdoor Attacks on Speech Recognition Models}

\author{Jinwen Xin \and Xixiang Lyu \and Jing Ma
\thanks{This is the authors' manuscript of a chapter published in Machine Learning for Cyber Security, Lecture Notes in Computer Science, vol. 13655, pp. 597--610 (2023). The final publication is available at \url{https://doi.org/10.1007/978-3-031-20096-0_45}.}}
\authorrunning{J. Xin et al.}
%
\institute{School of Cyber Engineering, Xidian University, Xi’an, China
\email{xxlv@mail.xidian.edu.cn}}
\maketitle              
%
\begin{abstract}
With the rapid development of deep learning, its vulnerability has gradually emerged in recent years. This work focuses on backdoor attacks on speech recognition systems. We adopt sounds that are ordinary in nature or in our daily life as triggers for natural backdoor attacks. We conduct experiments on two datasets and three models to validate the performance of natural backdoor attacks and explore the effects of poisoning rate, trigger duration and blend ratio on the performance of natural backdoor attacks. Our results show that natural backdoor attacks have a high attack success rate without compromising model performance on benign samples, even with short or low-amplitude triggers. It requires only 5\% of poisoned samples to achieve a near 100\% attack success rate. In addition, the backdoor will be automatically activated by the corresponding sound in nature, which is not easy to be detected and will bring severer harm.

\end{abstract}
\section{Introduction}
Over the past years, with the rapid development of deep learning, human beings have entered the era of AI. Deep learning systems have become prevalent in various fields, including face recognition \cite{parkhi2015deep,sun2014deep}, object detection \cite{redmon2016you}, machine translation \cite{wu2016google} and speech recognition \cite{yu2016automatic}. However, with the implementation of various deep learning applications, the vulnerability of deep neural networks (DNNs) has gradually emerged. For example, the adversary can fool the DNNs with adversarial examples \cite{akhtar2018threat,madry2017towards}.

Compared with adversarial attacks, which mainly occur in the inference stage of DNNs, backdoor attacks occur in the training stage, which needs large datasets and powerful computing resources. To reduce costs, users may adopt third-party datasets and third-party platforms, or directly adopt models provided by third parties \cite{li2022backdoor}. Meanwhile, transfer learning \cite{weiss2016survey} and federated learning \cite{bonawitz2019towards} are becoming more prevalent. In these untrustworthy scenarios, backdoor attacks can pose a huge threat to the security of DNNs. In general, backdoor attacks work by implanting a backdoor in DNNs during the training stage so that the infected DNNs perform well on benign samples, whereas their predictions will be maliciously changed if the buried backdoor is activated by the preset trigger \cite{li2022backdoor}. 

Speech recognition (SR) plays a key role in many fields, such as voice input, automatic drive and human-computer interaction. At present, most of the researches on backdoor attacks focus on the field of computer vision (CV), but few on SR. However, DNNs of SR are easily disturbed by the noise, which makes them more vulnerable to backdoor attacks. For some security-sensitive applications, such as voice commands in autonomous driving, the existence of backdoor will pose a huge threat. So backdoor attacks on SR deserve attention.

We focus on the design of triggers for backdoor attacks. To the best of our knowledge, there are mainly two types of triggers used in the existing backdoor attacks on SR. One is injecting random noise as a trigger into raw audio samples \cite{liu2017trojaning,xu2021detecting,tang2020embarrassingly}, and the other is injecting a sound wave of a certain frequency into the original audio \cite{zhai2021backdoor}, which can be an ultrasonic pulse that humans cannot hear \cite{koffas2021can}. Almost all existing backdoor attack methods for SR adopt random or meaningless noise as triggers. In order to activate the backdoor in the inference stage, the above works require the adversary to play the sound wave corresponding to the trigger, which undoubtedly increases the risk of being detected.

\begin{figure}
\includegraphics[width=\textwidth]{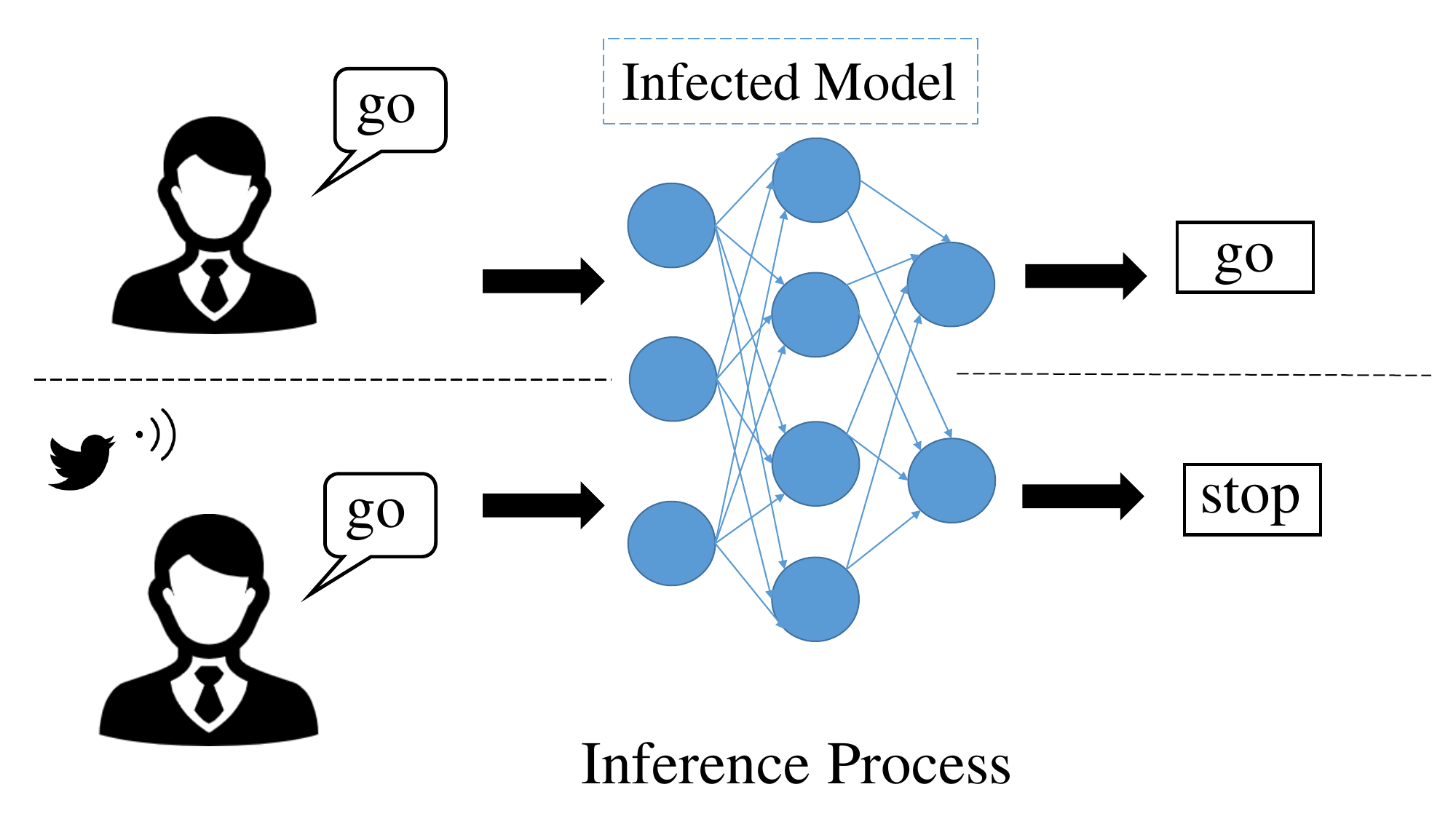}
\caption{A natural backdoor attack on SR models.} \label{fig1}
\end{figure}

In this work, we propose to adopt sounds that exist in nature or in our daily life as natural triggers, such as sound of rain, whistle and bird call, to implement a natural backdoor attack on SR models (Fig.~\ref{fig1}). Compared with the current methods, our method has the following advantages: Firstly, triggers are hidden. Sounds in nature or in our daily life do not attract human attention because these sounds are ordinary. And our experiments show that few poisoning samples can achieve a high attack success rate, so natural backdoor attacks are covert. Secondly, it is easier to activate the backdoor in SR models during the inference phase. Our method can cause the backdoor to be automatically activated by sounds in nature or in our daily life, which poses a greater threat to SR systems. Our main contributions are as follows:
\begin{enumerate} 
        \item We propose to adopt sounds that exist in nature or in our daily life as natural triggers to implement a natural backdoor attack on SR models, which is stealthy and harmful.
        \item We verify that natural backdoor attacks are still effective in real physical scenarios and are suitable for Clean-label attacks.
        \item We evaluate the performance of backdoor attacks with natural triggers and explore the influence of various factors on natural backdoor attacks.
\end{enumerate}

The rest of this paper is organized as follows: Sect.~\ref{section:A} provides the existing works related to backdoor attacks. Implementation method of natural backdoor attacks is presented in Sect.~\ref{section:B}. The experimental setup and results are demonstrated in Sect.~\ref{section:C}. Our summary is arranged in Sect.~\ref{section:D}.

\section{Related Work}
\label{section:A}
The concept of the backdoor attack against DNNs was first proposed by Gu et al. \cite{gu2019badnets}. Recently, many works have been proposed to implement the backdoor attack on image domain \cite{gu2019badnets,chen2017targeted,liu2020reflection}. Liu et al. utilize the natural phenomena (i.e., the reflection) common in life as the trigger to make the attack method stealthy \cite{liu2020reflection}. Meanwhile, backdoor attacks on natural language processing (NLP) are also explored in \cite{qi2021mind,qi2021hidden}. Qi et al. proposed adopting the syntactic structure as the trigger in textual backdoor attacks, in which trigger-embedded samples are not easily detected \cite{qi2021hidden}.

Compared to the above works of backdoor attacks on image and NLP domains, there are relatively few works on speech recognition (SR). Initially, some studies tried to apply backdoor attack methods suitable for CV to SR \cite{liu2017trojaning,xu2021detecting,tang2020embarrassingly}. Liu et al. implanted a backdoor in the SR model by injecting background noise to the raw audio samples and retrained the model to recognize the poisoned audio as an adversary-specified word \cite{liu2017trojaning}. Xu et al. conducted a backdoor attack on SR system by generating a random sound signal as a trigger \cite{xu2021detecting}. Tang et al. directly stamped the trigger pattern which was initially designed for image recognition on the spectrogram of the raw audio sample \cite{tang2020embarrassingly}. Experiments in the above works have proved that SR models are also vulnerable to backdoor attacks. However, these studies are not specific to SR. For example, the method of directly stamping the trigger pattern on the spectrogram of raw audio samples is difficult to implement attacks in the inference stage. Subsequently, some backdoor attack methods for SR were proposed. Koffas et al. implemented a backdoor attack on SR systems using an inaudible trigger which was an ultrasonic pulse \cite{koffas2021can}. Zhai et al. designed a clustering-based attack scheme to implement a backdoor attack on speaker verification models \cite{zhai2021backdoor}. Ye et al. proposed a dynamic backdoor attack method against SR models, named DriNet \cite{ye2022drinet}. In order to activate the backdoor in the inference stage, the above works require the adversary to play the sound wave corresponding to the trigger, which undoubtedly increases the risk of being detected. 

In contrast, our method can cause the backdoor to be automatically activated by sounds in nature or in our daily life, which poses serious threats to SR systems. Moreover, similar to random noise and the ultrasonic pulse, which are not noticed or even heard by humans, sounds in nature or in our daily life do not attract human attention as these sounds are ordinary, so our method is covert.

\section{Methodology}
\label{section:B}
\subsubsection{Threat Model.} The attacker follows the grey box setting: has no knowledge about the model architecture, parameters and training process of the DNNs, but can control a small number of training samples. This kind of threat exists in real scenarios, speech recognition systems often adopt the dataset collected in the form of crowdsourcing, so that malicious participants can upload malicious data to implement attacks.
\subsubsection{Attack Target.} In general, the attacker adopts data poisoning to generate an infected model. It will be predicted as the ground-truth label for benign voice input, but for voice input mixed with the preset trigger, it will be predicted as the attacker-specified label.

\subsection{Poisoning-based Backdoor Attacks on Speech Recognition}
\subsubsection{DNNs for Speech Recognition.}In general, define $D_{train}=\{(x_i,y_i)\}_{i=1}^N$ as an original training dataset for a classification task, which contain $N$ benign audio samples, ${x_i}\in{X}$ and ${y_i}\in{Y}=\{1,2,\cdots ,K\}$, $y_i$ is the ground-truth label of the input $x_i$. The target of DNNs for SR is to learn a benign model $F_\omega :X\rightarrow{Y}$ where $X$ is the input space and $Y$ is the label space. The purpose of model training is to find the optimal parameter $\omega$  to minimize the distance between the output predicted by the model $F_{\omega}$ and the ground-truth labels, distance is usually measured with a loss function $\mathcal{L}$. The specific calculation formula is as follows:
\begin{equation}
\omega ^*=\mathop{\arg\min}\limits_{\omega}\sum_{i=1}^N \mathcal{L}(F_\omega (x_i),y_i)
\end{equation}

\subsubsection{Formulation of Data Poisoning.}
In the poisoning-based backdoor attack, the poisoned samples dataset is generated by revising part of samples from the original training dataset: $D_{poison}=\{(G_t(x_i),y^*)\}_{i=1}^P$, where $G_t:X\rightarrow X$ indicates the attacker-specified poisoned audio sample generator with the trigger audio $t$, $y^*$ is the attacker-specified target label. Then the poisoned training set $D_{train}^*=D_{train}\bigcup D_{poison}$ is used to train an infected model $F_\lambda ^*$. For a test dataset $D_{test}=\{(x_i,y_i)\}_{i=1}^M$, infected model will correctly predict the benign test samples: $F_\lambda^*(x_t)=y_t$, but would classify the trigger-embedded inputs as the target label: $F_\lambda^*(G_t(x_i))=y^* $. Let $\frac{P}{N}$ indicates the poisoning rate.

\subsubsection{Measure Metrics.}
To measure the performance of backdoor attacks, two common metrics are introduced \cite{li2022backdoor}: (1) Benign Accuracy (BA), which indicates the prediction accuracy of the infected model on benign test samples; and (2) Attack Success Rate (ASR), which indicates the proportion of trigger-embedded samples that are successfully predicted as the target label by the infected model. In general, higher ASR and BA mean better performance of backdoor attacks. Besides, a lower poisoning rate and unnoticed triggers are needed to make attacks stealthier.

\subsection{Backdoor Attacks with Natural Trigger}

\subsubsection{Generation of Natural Triggers.}
We adopt sounds that are ordinary in nature or in our daily life as triggers. For sounds in nature, there are sound of rain, thunder, bird call and so on. For sounds in daily life, we can choose whistle, ringtone, etc. Sounds with a high probability of occurrence are easy to activate the backdoor, but are easier to detect, such as rain, whistle. Sounds with a low probability of occurrence are converse, such as thunder. We directly adopt the open source data to obtain natural trigger audio. The sampling rate of the trigger audio is the same as the original audio samples.

\subsubsection{Embedding of Triggers.}
We adopt the method of Time Domain Synthesis Strategy: The audio file is loaded to get the waveform, which is a one-dimensional vector. $x=\{a_1,a_2,a_3\cdots a_{l_1}\}$ is defined as the waveform of the original audio sample, and $\delta =\{b_1,b_2,b_3\cdots b_{l_2}\}$ is defined as the waveform of the trigger audio. $l_1$ and $l_2$ represent the length of $x$ and $\delta$, respectively, $l_1\geq l_2$. $G_\delta  (x)=x^*$, define $x^*$ as the trigger-embedded inputs. The specific calculation process is as follows.

\begin{figure}
\includegraphics[width=\textwidth]{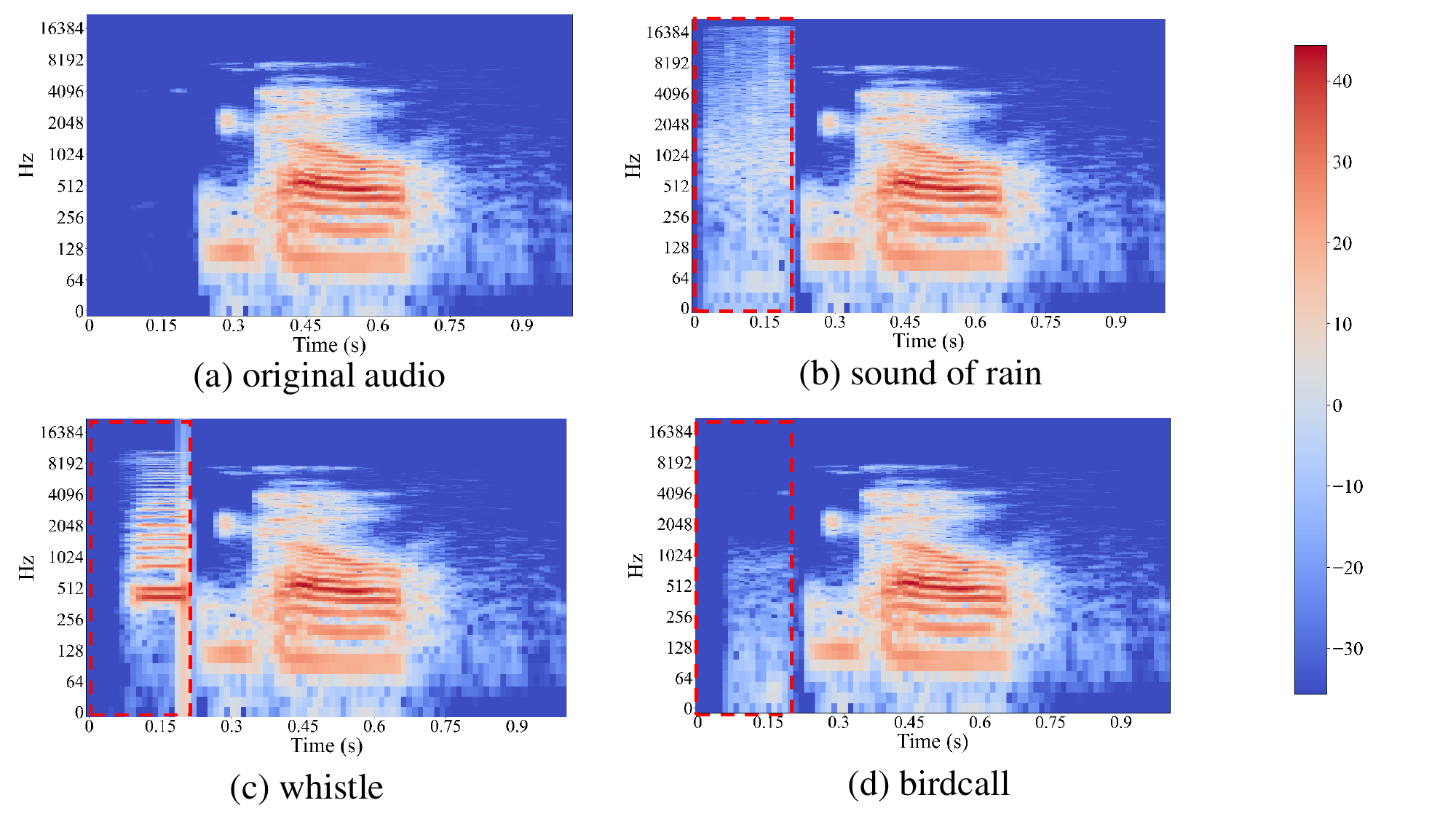}
\caption{Poisoned audio samples with different natural triggers: (a) original audio sample; (b) natural trigger of sound of rain; (c) natural trigger of whistle; (d) natural trigger of bird call.} \label{fig2}
\end{figure}

\begin{algorithm} 
\caption{Calculation process: $G_\delta (x)=x^*$} 
\label{alg1} 
\begin{algorithmic}[1] 
\REQUIRE Variables: $x,\delta,l_1,l_2$ 
\ENSURE Variables: $x^*$ 
\FOR{$i=1$ to $l_1$}
\IF{$i\leq l_2$}
\STATE $x_i^*\leftarrow x_i+\delta _i$
\ELSE
\STATE $x_i^*\leftarrow x_i$
\ENDIF
\ENDFOR
\end{algorithmic}
\end{algorithm}
The poisoning-based backdoor attack is the most commonly adopted attack method in backdoor attacks. We adopt the poisoning-based natural backdoor attack scheme that requires three steps to implement the attack. (1) Select sounds that are ordinary in nature or in our daily life as triggers, such as sound of rain, whistle and bird call, which can be obtained from open-source audio. (2) To generate poisoning samples, we add natural triggers to the original audio samples. Here we adopt the Time Domain Synthesis Strategy. (3) Add the poisoning samples to the training dataset to train an infected model. The spectrogram of the poisoned audio samples with different natural triggers is shown in Fig.~\ref{fig2}. Here we set the trigger duration to 0.2s and the original audio duration to 1s.

\section{Experiments of Natural Backdoor Attacks}
\label{section:C}
\subsection{Experimental Setup}
\subsubsection{Baseline Datasets and Models.} We conduct experiments on two datasets for speech classification tasks. One is Speech Commands Dataset Version 2 (SCDv2) \cite{warden2018speech}, also used as a benchmark dataset in other works \cite{koffas2021can,ye2022drinet,samizade2020adversarial}. Source audio adopts 44.1 kHz sampling rate, we selected 10 keywords in this dataset to form a 10-classes task. Since DNNs require inputs of consistent length, after discarding audio which are less than 1 second, we get 22384 audio samples. The other is Eating Sound Collection (ESC) \cite{kaggle20}, which is a 20-classes task to identify the sounds of eating different foods. Source audio use 16 kHz sampling rate. Similarly, we discard audio samples which are less than 3 seconds. For ESC, we adopt mini-CNN which is used as the baseline model in Ali Tianchi Competition \cite{tianchi21}. For SCDv2, we use CNN and LSTM. The CNN was also adopted in \cite{koffas2021can,samizade2020adversarial}, and the LSTM was introduced in \cite{de2018neural}. In addition, we set the epoch to 300 and the learning rate to 0.0001.

\subsubsection{Feature Extraction.} 
\begin{figure}
\includegraphics[width=\textwidth]{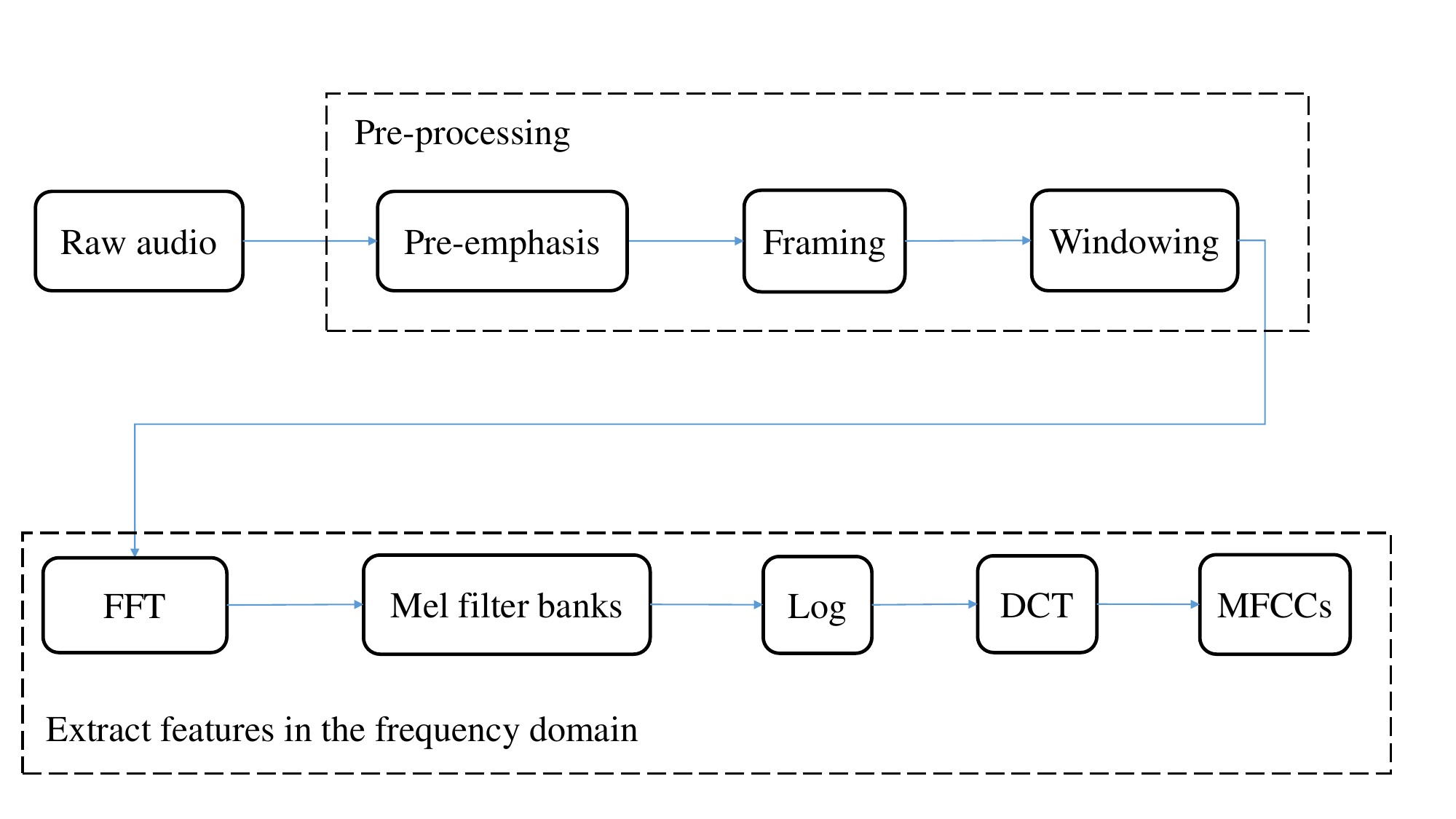}
\caption{Extract MFCCs as features from raw audio.} \label{fig3}
\end{figure}

In the training stage of DNNs, the first step is to extract features from the original input samples. Mel frequency cepstrum coefficient (MFCC) is one of the most commonly used features in SR. The specific process of extracting MFCC is shown in Fig.~\ref{fig3}. Firstly, preprocess the raw audio, including pre-emphasis, framing and windowing. After that, the frequency domain features are obtained after the fast Fourier transform (FFT). After the Mel filter banks, logarithmic and discrete cosine transform (DCT), we get MFCCs which are the input of DNNs.

\subsection{Evaluation of Natural Backdoor Attacks}
\label{section:D2}
We adopt three sounds that are ordinary in nature or in our daily life as natural triggers, including the sound of rain, whistle and bird call. In addition, we add random noise and ultrasound whose frequency is 21 kHz in related work \cite{liu2017trojaning,koffas2021can} as comparative experiments. 
We utilize the above five triggers to conduct the poisoning-based backdoor attack. Here we set the duration of the trigger to be the same as the duration of the original audio, the poisoning rate to 5\%. And we employ ASR and BA to measure the performance of backdoor attacks with different triggers. The results are shown in Table~\ref{tab1}. ACC represents the prediction accuracy of the test dataset on the benign model.

\begin{table}
\centering
\caption{Evaluation of natural backdoor attacks.}\label{tab1}
\begin{tabular}{@{}|l|l|l|l|l|l|l|@{}}
\toprule
Dataset & Model & Type                      & Trigger       & ACC(\%) & BA(\%) & ASR(\%) \\ \midrule
\multirow{5}{*}{ESC}    & \multirow{5}{*}{mini-CNN} & \multirow{2}{*}{Related work} & Random noise & \multirow{5}{*}{97.32} & 96.70 & 100.00 \\ \cmidrule(lr){4-4} \cmidrule(l){6-7} 
        &       &                           & Ultrasound    &         & 96.81  & 99.79   \\ \cmidrule(lr){3-4} \cmidrule(l){6-7} 
        &       & \multirow{3}{*}{Our work} & Sound of rain &         & \textbf{96.75}  & \textbf{99.79}   \\ \cmidrule(lr){4-4} \cmidrule(l){6-7} 
        &       &                           & Whistle       &         & \textbf{97.53}  & \textbf{99.19}   \\ \cmidrule(lr){4-4} \cmidrule(l){6-7} 
        &       &                           & Bird call      &         & \textbf{96.96}  & \textbf{96.78}   \\ \midrule
\multirow{10}{*}{SCDv2} & \multirow{5}{*}{CNN}      & \multirow{2}{*}{Related work} & Random noise & \multirow{5}{*}{92.82} & 92.68 & 100.00 \\ \cmidrule(lr){4-4} \cmidrule(l){6-7} 
        &       &                           & Ultrasound    &         & 90.88  & 14.58   \\ \cmidrule(lr){3-4} \cmidrule(l){6-7} 
        &       & \multirow{3}{*}{Our work} & Sound of rain &         & \textbf{92.24}  & \textbf{99.22}   \\ \cmidrule(lr){4-4} \cmidrule(l){6-7} 
        &       &                           & Whistle       &         & \textbf{93.02}  & \textbf{99.94}   \\ \cmidrule(lr){4-4} \cmidrule(l){6-7} 
        &       &                           & Bird call      &         & \textbf{92.95}  & \textbf{99.38}   \\ \cmidrule(l){2-7} 
                        & \multirow{5}{*}{LSTM}     & \multirow{2}{*}{Related work} & Random noise & \multirow{5}{*}{92.94} & 89.98 & 99.98  \\ \cmidrule(lr){4-4} \cmidrule(l){6-7} 
        &       &                           & Ultrasound    &         & 89.54  & 2.45    \\ \cmidrule(lr){3-4} \cmidrule(l){6-7} 
        &       & \multirow{3}{*}{Our work} & Sound of rain &         & \textbf{90.78}  & \textbf{97.74}   \\ \cmidrule(lr){4-4} \cmidrule(l){6-7} 
        &       &                           & Whistle       &         & \textbf{92.42}  & \textbf{99.97}   \\ \cmidrule(lr){4-4} \cmidrule(l){6-7} 
        &       &                           & Bird call      &         & \textbf{91.18}  & \textbf{96.13}   \\ \bottomrule
\end{tabular}
\end{table}

For ESC, backdoor attacks with five different triggers all have high ASR, and BA does not change much relative to ACC. For SCDv2, since the sampling rate of the dataset is 16kHz, according to the Channon Nyquist sampling theorem, an ultrasonic pulse will suffer loss in this sampling rate, so ASR is very low. Natural triggers have good performance, and the ASR for the CNN model has reached more than 99\%. For the LSTM model, ASR can also reach more than 96\%.

According to the experimental results, natural backdoor attacks have a high ASR without compromising model performance on benign samples. Compared with related works, our scheme has no restriction on the sampling rate of raw audio samples. Considering that the SCDv2 dataset is a more complicated speech recognition task compared to ESC, the following experiments are all for SCDv2. We choose CNN and LSTM as the infected models and the sound of rain as the natural trigger.

\begin{figure}
\includegraphics[width=\textwidth]{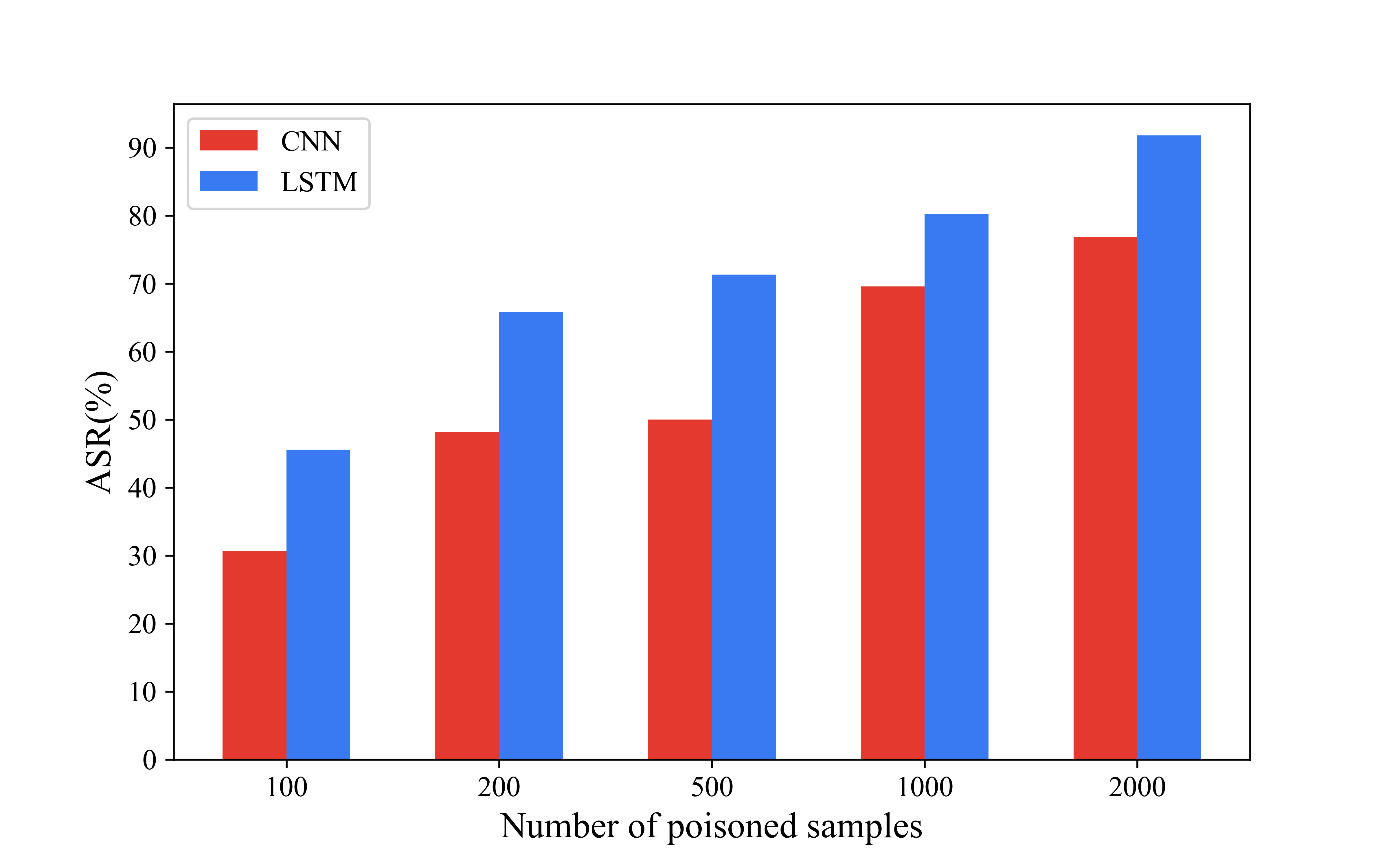}
\caption{A natural backdoor attack in real physical scenarios.} \label{fig4}
\end{figure}

In addition, we also verify the effectiveness of our method in real physical scenarios where we use the sound of cicadas in summer as the natural trigger. We recorded human voices accompanied by sound of cicadas as poisoned training samples. The results of our experiments are shown in Fig.~\ref{fig4}. The experimental results show that our method can still maintain a high ASR in real physical scenarios. The buried backdoor can be triggered by the sound in natural, which poses a challenge to the security of speech recognition systems.

\begin{figure}
\includegraphics[width=\textwidth]{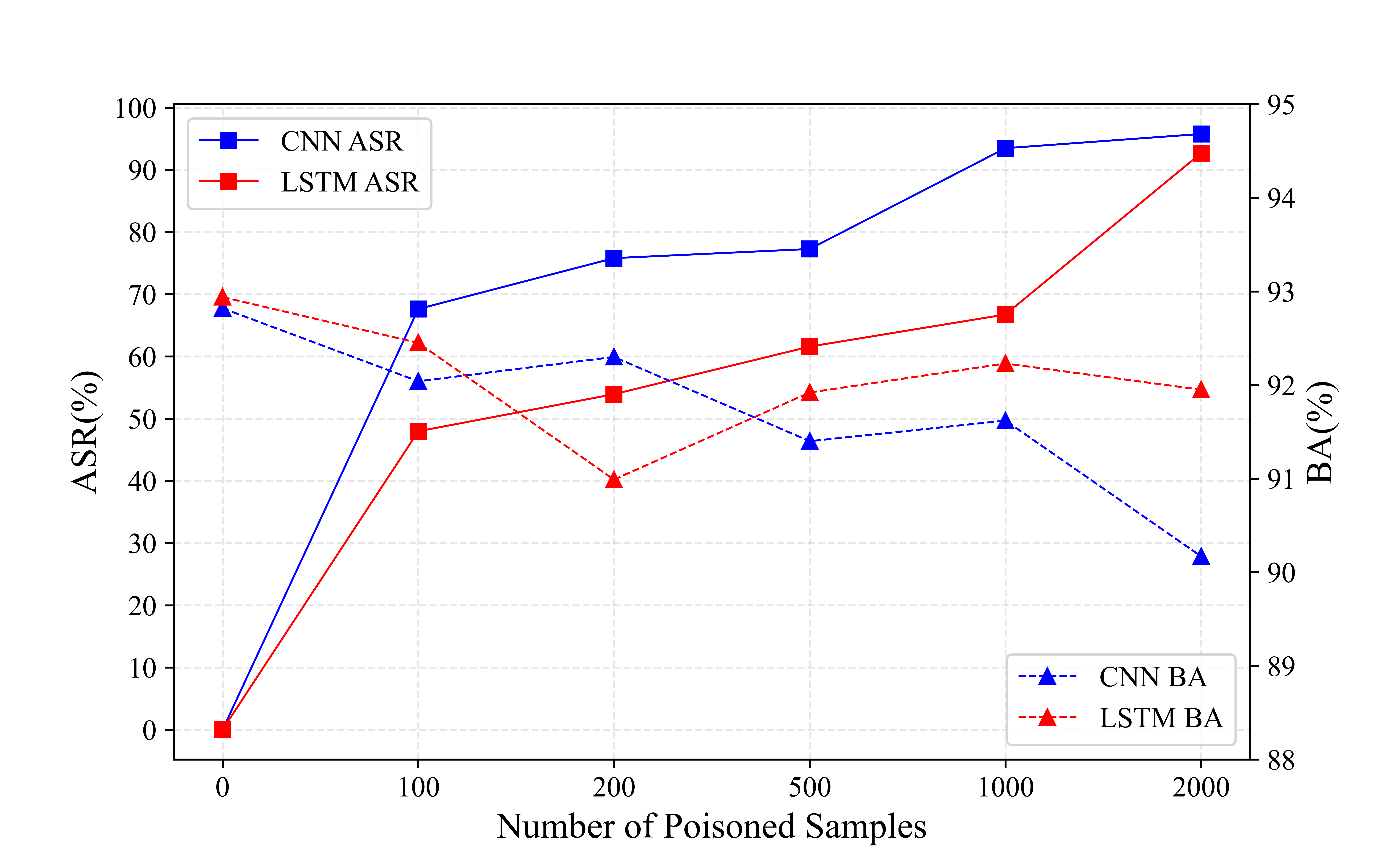}
\caption{Clean-label attacks with different numbers of poisoned samples.} \label{fig8}
\end{figure}

In our above experiments, the generation of poisoned samples requires modifying the labels of the samples, which makes the poisoned samples easily detected upon human inspection. Our attack method can also be extended to Clean-label attacks \cite{turner2018clean}, which means the adversary can insert a trigger without modifying the label of the samples, thus making our attack more invisible. Our experimental results are shown in Fig.~\ref{fig8}. Compared with Poison-label attacks, Clean-label attacks require more poisoned samples. When the poisoning rate reaches 5\% (1000 poisoned samples), the ASR of the backdoor attack on CNN reaches more than 90\%. Clean-label attacks on LSTM models require a higher poisoning rate. 

\subsection{Effect of Different Factors on ASR}
We examine the impact of different factors on the performance of natural backdoor attacks, including poisoning rate, trigger duration and blend ratio. 

\subsubsection{Poisoning Rate.}
\begin{figure}
\includegraphics[width=\textwidth]{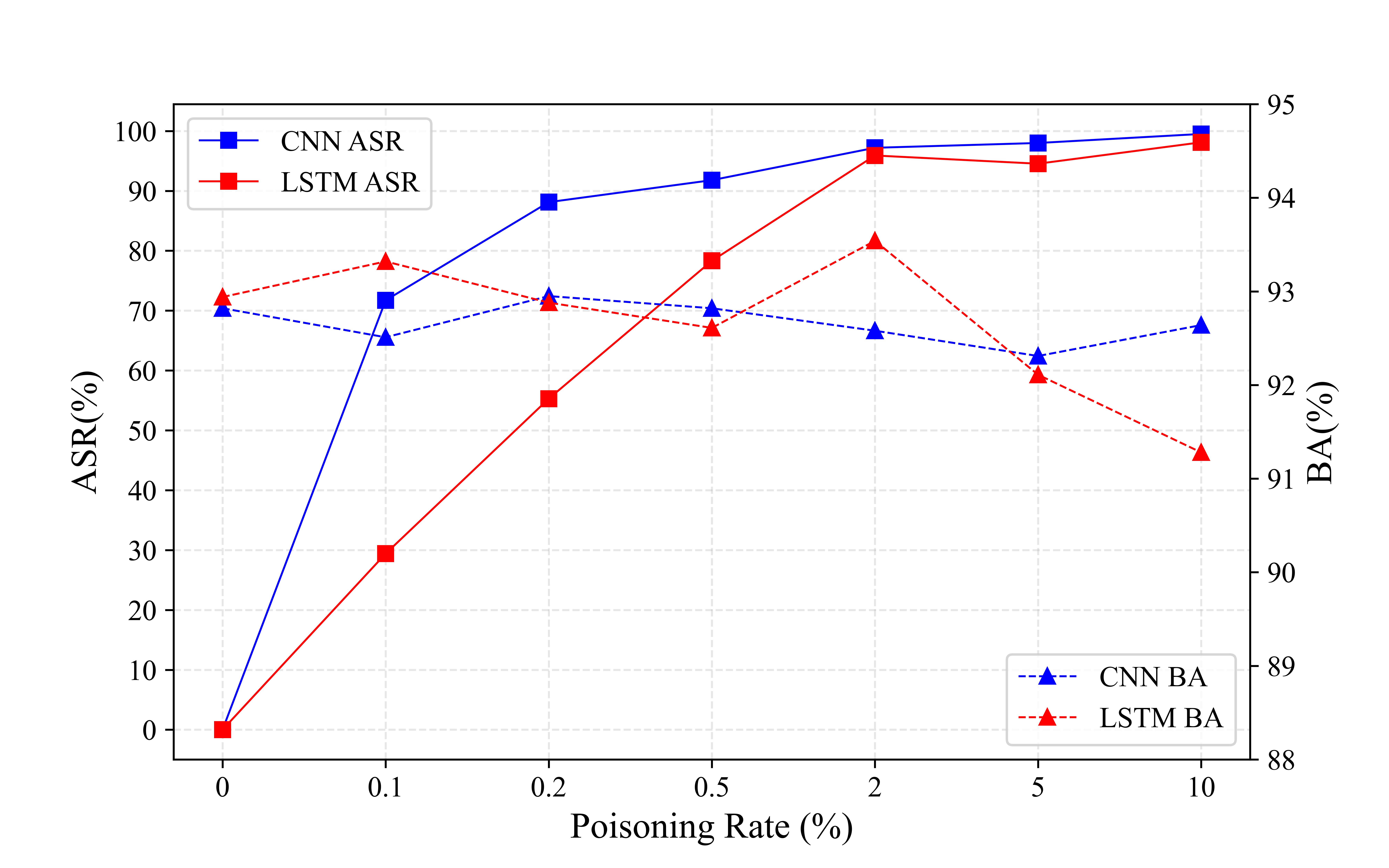}
\caption{The performance of natural backdoor attacks under different poisoning rates.} \label{fig5}
\end{figure}

We explore the performance of natural backdoor attacks under different poisoning rates. As shown in Fig.\ref{fig5}, as the poisoning rate increases, the ASR continues to in-crease, and there are only slight fluctuations for BA. When the poisoning rate reaches 2\%, the ASR of the backdoor attack on both CNN and LSTM models reaches more than 90\%. When the poisoning rate reaches 5\%, ASR is close to 100\%. Compared to CNN, implementing attacks on LSTM requires a larger poisoning rate.

\subsubsection{Trigger Duration.}
\begin{figure}
\includegraphics[width=\textwidth]{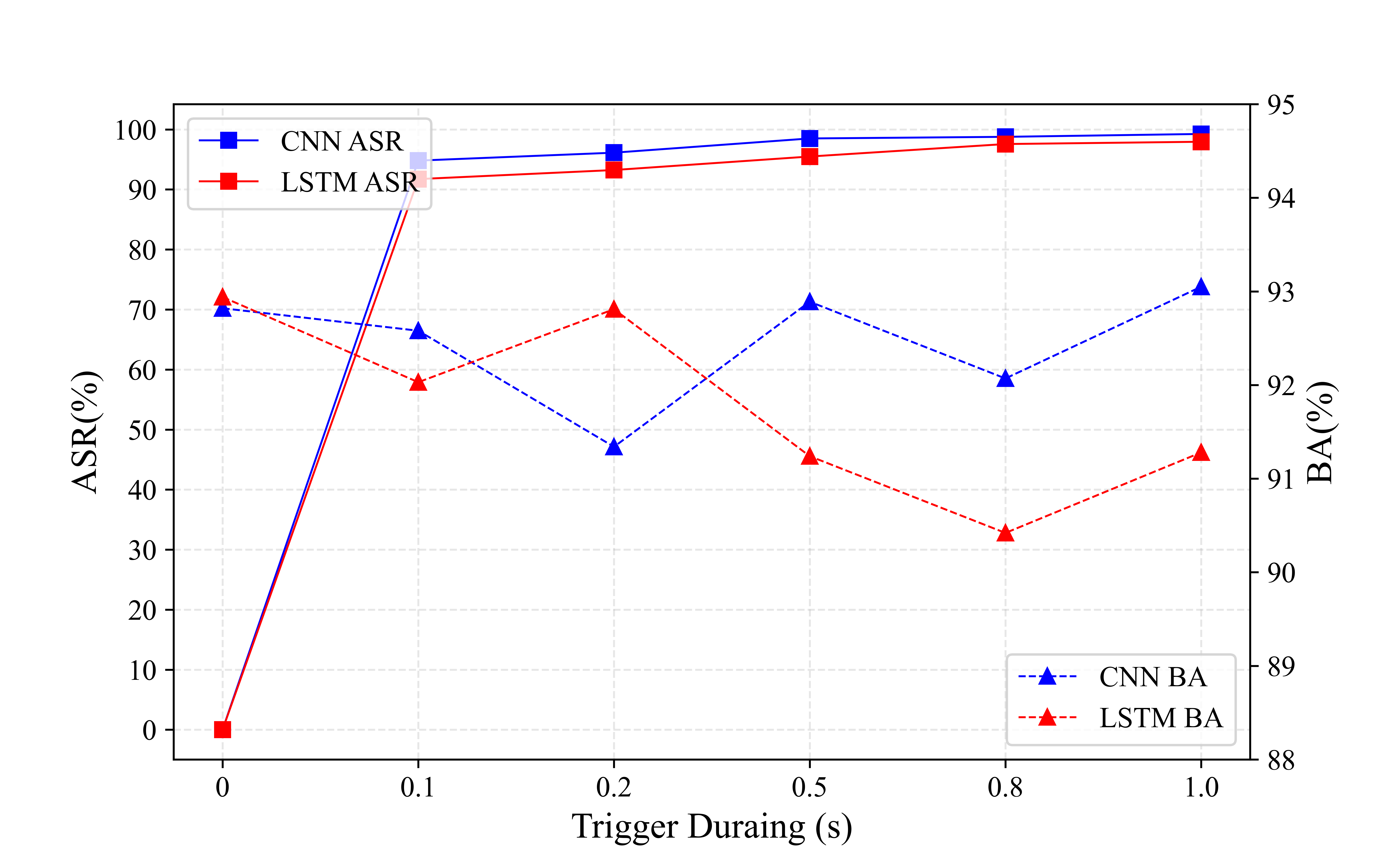}
\caption{The performance of natural backdoor attacks under different trigger duration.} \label{fig6}
\end{figure}

In Sect.~\ref{section:D2}, the duration of the trigger we set is the same as the audio samples, which is 1 second. Here we explore the effect of natural trigger duration on the performance of natural backdoor attacks. The results are shown in Fig.~\ref{fig6}. As the trigger duration increases, the ASR continues to increase, BA has only minor changes. And the change curve of the ASR for the two infected models remains consistent. When the trigger duration reaches 0.1 s, the ASR exceeds 90\%. When it reaches 0.8 s, the ASR is close to 100\%.

\subsubsection{Blend Ratio.}
\begin{figure}
\includegraphics[width=\textwidth]{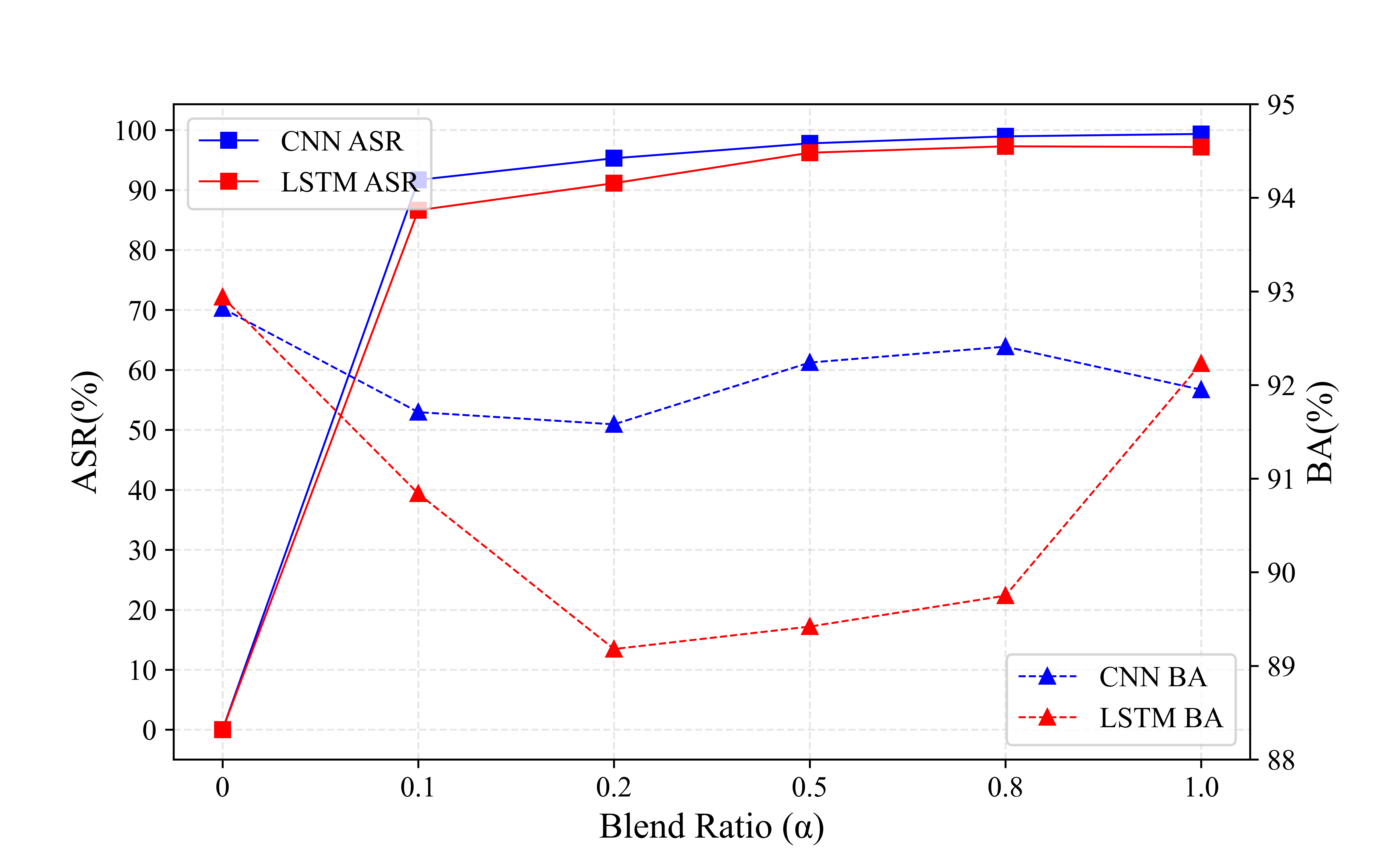}
\caption{The impact of different blend ratios on natural backdoor attacks.} \label{fig7}
\end{figure}

We employ the Blended Injection Strategy \cite{chen2017targeted} to generate poisoned samples. The parameters $s$ and $t$ represent the one-dimensional vector representation of the original audio sample and the trigger signal, respectively. The parameter $\alpha$ represents the blend ratio. The poisoned samples are generated as follows: 
\begin{equation}
\prod _\alpha ^{blend}(s,t)=s+\alpha \cdot t 
\end{equation}
We explore the impact of different blend ratios on natural backdoor attacks, the result is shown in Fig.~\ref{fig7}. As the blend ratio increases, the ASR continues to increase, and BA has only minor changes. And the change curve of the ASR for the two infected models remains consistent. When the blend ratio reaches 0.1, the ASR exceeds 85\%. When it reaches 0.8, the ASR is close to 100\%.

\section{Conclusion}
\label{section:D}
In this paper, we propose natural backdoor attacks on speech recognition models using sounds which are ordinary in nature or in our daily life as natural triggers. Our results show that natural backdoor attacks have a high attack success rate without compromising model performance on benign samples, even with short or low-amplitude triggers. Only 5\% of the poisoned samples are needed to achieve a near 100\% attack success rate. In addition, natural backdoor attacks are still effective in real physical scenarios and are suitable for Clean-label attacks. And the backdoor can be automatically activated by the corresponding sound in nature as a trigger, which will bring severer harm.

%
%

%
\bibliographystyle{splncs04}
\bibliography{mybibliography}
\end{document}